\documentclass[twocolumn,showpacs,prd]{revtex4}
\usepackage{graphicx}
\usepackage{dcolumn}
\usepackage{bm}
\begin{document}

\title
{On the Origins of the Planck Zero Point Energy in Relativistic Quantum Field Theory }
\author{A. Widom and J. Swain}
\affiliation{Physics Department, Northeastern University, Boston MA USA}
\author{Y.N. Srivastava}
\affiliation{Physics Department, University of Perugia, Perugia IT}

\begin{abstract}

It is argued that the zero point energy in quantum field theory 
is a reflection of the particle anti-particle content of the theory. This 
essential physical content is somewhat disguised in electromagnetic theory 
wherein the photon is its own anti-particle. To illustrate this point, we 
consider the case of a charged Boson theory $(\pi^+,\pi^-)$ wherein the 
particle and anti-particle can be distinguished by the charge $\pm e$. 
Starting from the zero point energy, we derive the Boson pair production 
rate per unit time per unit volume from the vacuum in a uniform external 
electric field. The result is further generalized for arbitrary spin $s$.  

\end{abstract}


\maketitle

\section{Introduction \label{intro}}

The zero point energy in Boson quantum field theories  
are a consequence of the Boson and anti-Boson content of the theory. This 
physical conclusion is masked in electromagnetic theory 
because the photon is its own anti-particle. If one considers a charged 
Boson theory, say \begin{math} (\pi^+,\pi^-) \end{math}, the 
particle and anti-particle can be distinguished in virtue of the electric 
charge. From the zero point energy, one may derive the known Boson pair 
production rate per unit time per unit volume from the vacuum in a uniform 
external electric field. The external electric field does the work 
required to excite the oscillators.

In Sec.\ref{pd} it is argued that the particle anti-particle content of 
the original Planck photon number distribution can be employed to explain 
the {\em zero point energy} level \begin{math} \hbar |\omega |/2 \end{math}.  
Relating the zero point energy to the particle content of field theory 
is not so evident for the case of photons because the photon and the 
anti-photon are the same particle.
In Sec. \ref{pion}, we consider spin zero Bosons that are charged, e.g. 
the \begin{math} (\pi^+,\pi^-) \end{math} system wherein the particle has 
charge \begin{math} +e \end{math} and the anti-particle has charge 
\begin{math} -e \end{math}. The application of a uniform electric field 
excites the charged Boson oscillators ripping  
\begin{math} (\pi^+,\pi^-) \end{math} pairs out of the vacuum. The external 
electric fields can cause the dielectric breakdown of the Boson vacuum. 
The rate of Boson pair production per unit time per unit volume is 
also computed here.
The case of Fermion pair production is
considered in Sec. \ref{cpp} together with a general formula for
production of charged particle-antiparticle pairs with arbitrary spin.
Concluding remarks are given in Sec. \ref{conc}.

\section{Planck Distributions \label{pd}}

Planck originally\cite{Planck:1901} discussed the mean 
number of photons of frequency \begin{math} \omega \end{math} in the 
thermal vacuum 
\begin{equation}
\bar{n}=\left[\frac{1}{e^{\hbar \omega /k_BT}-1}\right]. 
\label{pd1}
\end{equation}
The mean thermal energy of an electromagnetic oscillator was thereby 
taken to be 
\begin{equation}
\bar{E}(\omega )=\hbar \omega \bar{n}
=\left[\frac{\hbar \omega }{e^{\hbar \omega /k_BT}-1}\right]. 
\label{pd2}
\end{equation}
Later\cite{Planck:1912} Planck somewhat arbitrarily added the zero-point energy
via a second theory, now assuming that absorption should be treated classically, but emission
by discrete quanta - a procedure which might seem odd by today's
standards, but this was at the very beginning of quantum theory. 
Einstein and Stern\cite{Einstein:1913} noted that the excess energy over 
and above the equipartition value obeyed 
\begin{equation}
\lim_{T\to \infty}
\left[\bar{E}(\omega )+\frac{\hbar \omega}{2}-k_BT\right]=0  
\label{pd2a}
\end{equation}
that might theoretically be regarded as slight evidence of a zero temperature 
energy of \begin{math}   \hbar \omega /2   \end{math}. A nice discussion
of zero point energy in early quantum theory can be found in the book by Milonni\cite{Milonni:1994}.

On the other hand, by actually solving the quantum mechanical 
simple harmonic oscillator one obtains a zero point energy term 
\begin{equation}
E_T(\omega )=\hbar \omega \left[\bar{n}+\frac{1}{2}\right] 
= \left(\frac{\hbar \omega }{2}\right)
\coth \left(\frac{\hbar \omega }{2k_BT}\right). 
\label{pd3}
\end{equation}
Let us note the following \\  
{\bf Theorem:}  
\begin{equation}
E_T(\omega )=\frac{1}{2}\left[\bar{E}(\omega )+\bar{E}(-\omega )\right]. 
\label{pd4}
\end{equation}
{\bf Proof:} Eq.(\ref{pd4}) is true in virtue of Eqs.(\ref{pd2})  
and (\ref{pd3}). Also note the zero point energy 
\begin{equation}
E_0(\omega ) \equiv \lim_{T\to 0^+} E_T(\omega )
=\left(\frac{\hbar |\omega |}{2}\right). 
\label{pd5}
\end{equation}
If one expresses this result in terms of the photon creation operator 
\begin{math} a^\dagger \end{math} and destruction operator 
\begin{math} a \end{math} with 
\begin{math} \big[a,a^\dagger \big]=1 \end{math}, then the photon 
number operator \begin{math} n=a^\dagger a  \end{math} enters into the 
Hamiltonian via the symmetrized product 
\begin{math} \big(a a^\dagger + a^\dagger a \big) \end{math} as   
\begin{equation}
{\cal H}=\frac{\hbar |\omega |}{2}\left(a^\dagger a + a a^\dagger \right) 
=\hbar |\omega |\left(n+\frac{1}{2}\right).  
\label{pd6}
\end{equation}
Eq.(\ref{pd6}) leads directly to Eq.(\ref{pd5}). 

Let us consider the physical meaning of Eq.(\ref{pd4}). The positive 
frequency \begin{math} \omega > 0  \end{math} represents a particle 
moving forward in time, while the negative frequency 
\begin{math} \omega < 0  \end{math} represents an anti-particle moving 
backward in time. Since the photon is its own anti-particle, the physical 
meaning of Eqs.(\ref{pd4}) and (\ref{pd5}) may be somewhat obscured. In 
order to make the particle content in the zero point energy more evident, 
let us consider a case wherein the particle and anti-particle are distinct. 

Interestingly, the simple argument that both positive and negative signs for 
$\omega$ should be treated symmetrically would have been available
to Planck, and many others afterwards to obtain the zero point energy from
Planck's first formula by simply symmetrizing it to reflect this, but this seems
to have been missed.

\section{Charged Spinless Bosons  \label{pion}}

The energy of a spinless charged scalar Boson in a uniform magnetic 
field \begin{math} {\bf B}=(0,0,B)=(0,0,|{\bf B}|) \end{math} is 
given by\cite{Landau:1980}  
\begin{eqnarray}
n=0,1,2,3,4, \cdots \ , 
\nonumber \\ 
{\bf p}=(0,0,p)\ , 
\nonumber \\ 
\epsilon_\pm (n,p,B)=\pm c\sqrt{m^2c^2 + p^2  +(2n+1)|\hbar e{\bf B}|/c}\ , 
\label{pion1}
\end{eqnarray}
wherein \begin{math} n \end{math} is the label for the circular Landau orbit, 
the momentum along the magnetic field axis is \begin{math} p=\hbar k \end{math}  
and \begin{math} \kappa =(mc/\hbar )\end{math} is the mass in inverse length 
units. Thus 
\begin{eqnarray}
\omega(n,k,B)=c\sqrt{\kappa^2 + k^2  +(2n+1)|e{\bf B}/\hbar c|}\ .  
\label{pion2}
\end{eqnarray}
The zero point charged Boson oscillator energies per unit volume counting 
the particle and anti-particle separately in virtue of the different 
charge \begin{math} \pm e \end{math} is determined by  
\begin{eqnarray}
U_0(B)=2\times 
\left[\left(\frac{eB}{2\pi \hbar c}\right)\sum_{n=0}^\infty 
\int_{-\infty}^\infty \frac{dk}{2\pi} 
\left[\frac{\hbar \omega(n,k,B)}{2}\right]\right].  
\label{pion3}
\end{eqnarray}
This vacuum energy per unit volume in a magnetic field is clearly divergent 
so one must {\em regulate} and {\em renormalize}. After doing both processes,  
a finite vacuum Boson energy per unit volume in a magnetic field 
\begin{math} U(B) \end{math} arises. 

\subsection{Gamma Function Regulation \label{gfr}} 
    
The Gamma function is defined in the \begin{math} {\Re e}\{z\}> 0 \end{math} 
part of the complex plane as  
\begin{equation}
\Gamma (z)=\int_0^\infty t^z e^{-t} \left(\frac{dt}{t}\right) ,
\label{gfr1}
\end{equation}
from which we find for \begin{math} a>0 \end{math} the identity  
\begin{equation}
a^{-z}=
\frac{1}{\Gamma(z)}\int_0^\infty s^z e^{-as} \left(\frac{ds}{s}\right).
\label{gfr2}
\end{equation}
In other regimes, \begin{math} \Gamma (z) \end{math} is defined by analytic 
continuation. This analysis is often assisted by the identity 
\begin{equation}
\Gamma (1+z)=z\Gamma (z).
\label{gfr3}
\end{equation}
For example, by putting \begin{math} z=-(1/2) \end{math} in Eq.(\ref{gfr3}), 
one finds 
\begin{equation}
\Gamma \left(\frac{1}{2}\right)=\sqrt{\pi } \ \ \ \ \ \Rightarrow 
\ \ \ \ \ \Gamma \left(-\frac{1}{2}\right)=-2\sqrt{\pi}\ .
\label{gfr4}
\end{equation}
Eqs.(\ref{gfr2}) and (\ref{gfr4}) lead to a {\em formally divergent integral}  
\begin{equation}
\sqrt{a}=-\left(\frac{1}{2\sqrt{\pi }}\right)
\int_0^\infty e^{-as} \left(\frac{ds}{s^{3/2}}\right).
\label{gfr5}
\end{equation}
For those readers who find it strange in Eq.(\ref{gfr5}) to put a finite positive 
quantity equal to a negative infinite quantity, we invite the reader to prove 
the following \\ 
{\bf Theorem:} For any \begin{math} a>0 \end{math} and 
\begin{math} b>0 \end{math}
\begin{equation}
\sqrt{a}-\sqrt{b}= \left(\frac{1}{2\sqrt{\pi }}\right) \int_0^\infty 
\left[e^{-bs}-e^{-as}\right]\left(\frac{ds}{s^{3/2}}\right).
\label{gfr5c}
\end{equation}
Subtractions will made below. Eqs.(\ref{pion3}) and (\ref{gfr5}) yield 
\begin{eqnarray}
U_0(B)=-\left(\frac{eB}{8\pi^{5/2}}\right) 
\sum_{n=0}^\infty \int_{-\infty}^\infty  
dk \int_0^\infty \left(\frac{ds}{s^{3/2}}\right) \times 
\nonumber \\ 
\exp\left[-\kappa ^2 s -k^2 s -(2n+1)\left|\frac{eB}{\hbar c}\right|s\right],  
\nonumber \\ 
U_0(B)=-\left(\frac{\hbar c}{16\pi^2}\right)\int_0^\infty 
\frac{ds}{s^3}e^{-\kappa ^2s}
\left[\frac{(eBs/\hbar c)}{\sinh(eBs/\hbar c)}\right] 
\label{gfr6}
\end{eqnarray}
that is still divergent. Now let us subtract the vacuum zero point oscillations when 
the magnetic field is zero, 
\begin{equation}
\tilde{U}(B)=U_0(B) - U_0(0).
\label{gfr7}
\end{equation}
The zero point oscillation energy per unit volume due to vacuum 
particle anti-particle pairs, say \begin{math} (\pi^+, \pi^-) \end{math}, 
virtual magnetic moments are thereby 
\begin{eqnarray}
\tilde{U}(B)=\left(\frac{\hbar c}{16\pi^2}\right)\times  
\nonumber \\ 
\int_0^\infty \frac{ds}{s^3}\ e^{-\kappa ^2s} 
\left[1-\left(\frac{eBs/\hbar c}{\sinh(eBs/\hbar c)}\right)\right],
\label{gfr8}
\end{eqnarray}
which still is divergent but only as a logarithm at small distance squared 
as \begin{math} s\to 0^+  \end{math}. Once the divergences are only 
logarithmic one may pass from {\em regularization} to {\em renormalization}.  

\subsection{Charge Renormalization \label{cr}} 
    
In quantum electrodynamics one starts with charges and fields described for 
the problem at hand by \begin{math} e_0  \end{math} and 
\begin{math} B_0 \end{math}. Both the fields and charges have to be 
renormalized in such a way that \begin{math} e_0 B_0=eB \end{math} and 
thus divergent logarithms are buried. The physical fields are defined so that 
the normal vacuum magnetic energy density is 
\begin{math} |{\bf B}|^2/8\pi  \end{math}. This can all be realized by a charge 
renormalization subtraction in Eq.(\ref{gfr8}). Thus, for scalar Boson fields 
the vacuum energy density is given by 
\begin{eqnarray}
U(B)=\left(\frac{\hbar c}{16\pi^2}\right) 
\int_0^\infty \frac{ds}{s^3}\ e^{-\kappa ^2s} \times  
\nonumber \\  
\left[1-\frac{(eBs/\hbar c)}{\sinh(eBs/\hbar c)}- 
\frac{(eBs/\hbar c)^2}{6}\right].
\label{cr1}
\end{eqnarray}
Eq.(\ref{cr1}) is both finite and exact for the sum of zero point 
oscillations of charged Boson spin zero systems. All divergences 
have been buried. The vacuum Boson magnetization is thereby 
\begin{equation}
{\bf M}=-\left(\frac{\partial U}{\partial {\bf B}}\right).
\label{cr2}
\end{equation}
Let us now consider what happens in an external electric field. 

\subsection{Photon Fields \label{pf}}

The electromagnetic fields associated with a photon obey the vacuum Maxwell 
equations
\begin{eqnarray}
div{\bf E}=0, \ \ \ \ \  div{\bf B}=0, 
\nonumber \\ 
curl{\bf E}=-\frac{1}{c}\left(\frac{\partial {\bf B}}{\partial t}\right),  
\ \ \ \ \ \ 
curl{\bf B}=\frac{1}{c}\left(\frac{\partial {\bf E}}{\partial t}\right).
\label{pf1}
\end{eqnarray}
In terms of the complex vector field
\begin{eqnarray}
{\bf F}={\bf E}+i{\bf B}, 
\nonumber \\ 
div{\bf F}=0, 
\ \ \ \ i\left(\frac{\partial {\bf F}}{\partial t}\right) 
= c\ curl{\bf F}.
\label{pf2}
\end{eqnarray}
Eq.(\ref{pf2}) describes a space-time Schr\"odinger equation for a photon. 
The square of the vector wave function, 
\begin{equation}
{\bf F}\cdot {\bf F} = |{\bf E}|^2-|{\bf B}|^2+2i{\bf E\cdot B},
\label{pf3}
\end{equation}
determines the Lorentz scalar 
\begin{math} |{\bf E}|^2-|{\bf B}|^2 \end{math} 
and Lorentz pseudo-scalar \begin{math} {\bf E\cdot B} \end{math}.
To go from a pure external magnetic field to a pure external electric 
field one takes \begin{math} B^2 \to -E^2 \end{math}. 
This allows us to obtain the Boson pair production rate 
\begin{math} \Gamma \end{math} per unit time per unit volume.

\subsection{Boson Pair Production \label{bpp}}

The transition rate per unit time per unit volume in an external 
electric field may be computed from 
\begin{equation}
\Gamma = -\left(\frac{2}{\hbar }\right) 
{\Im m}\ U(B\to -iE).
\label{bpp1}
\end{equation}
Eqs.(\ref{cr1}) and (\ref{bpp1}) imply 
\begin{eqnarray}
\Gamma = \frac{c}{8\pi ^3}\left(\frac{eE}{\hbar c}\right)^2 \times 
\nonumber \\  
\sum_{n=1}^\infty \frac{(-1)^{n+1}}{n^2} 
\exp \left( - \pi n\left|\frac{m^2c^3}{\hbar e E}\right|\right). 
\label{bpp2} 
\end{eqnarray}
A uniform electric field can thereby excite the charged Boson oscillators 
emitting pairs \begin{math} (\pi^+, \pi^-) \end{math} from the vacuum. The 
electric field does the work required to {\em break down the vacuum}.   

\section{Charged Particle Paths \label{cpp}} 

Let us here consider how the zero point energy is expressed in terms of 
paths forward in time (particle) and backward in time (anti-particle). If 
a particle {\em over here} can go forward in time and then backward in time 
to {\em over there}, then {\em right now} the quantum amplitude is 
{\em partly over here} and {\em partly over there} which evidently 
{\em spreads out} the zero point ground state wave function. Let us 
turn the above physical picture into a computation of the charged 
pair production rate in Eq.(\ref{bpp2}). Let us at 
first work in one space and one time (1+1) dimensions. With a small 
modification, this leads to a correct description in physical three space and one time 
(3+1) dimensions. 

In (1+1) dimensions, the energy-momentum relation reads 
\begin{equation}
{\cal E}^2-c^2p^2=(mc^2)^2 .  
\label{cpp1} 
\end{equation}
Since energy is force times distance and momentum is force times time, 
Eq.(\ref{cpp1}) reads 
\begin{equation}
(eEx)^2-c^2(eEt)^2=(mc^2)^2 ,  
\label{cpp2} 
\end{equation}
or in terms of the particle acceleration \begin{math} a \end{math},   
\begin{equation}
a=(eE/m),  
\label{cpp3} 
\end{equation}
Eq.(\ref{cpp2}) reads 
\begin{equation}
x^2-c^2t^2=(c^2/a)^2   
\label{cpp4} 
\end{equation}
that describes classical paths. The particle path forward in time is 
\begin{equation}
x_+(t)=\sqrt{c^2t^2+(c^2/a)^2}    
\label{cpp5} 
\end{equation}
while the anti-particle path backward in time is 
\begin{equation}
x_-(t)=-\sqrt{c^2t^2+(c^2/a)^2}    
\label{cpp6} 
\end{equation}
Pair production at {\em time zero} requires a space-like transition 
from \begin{math} x_-(0)=-(c^2/a) \end{math} to 
\begin{math} x_+(0)=(c^2/a) \end{math} along the semicircle in Euclidean time 
\begin{math} t_{\cal E} \end{math}, i.e. Eq.(\ref{cpp4}) reads in Euclidean 
time 
\begin{equation}
x^2+c^2t_{\cal E}^2=(c^2/a)^2 .   
\label{cpp7} 
\end{equation}
The arc length of the semicircle is \begin{math} s=\pi (c^2/a) \end{math} 
giving rise to the Euclidean action 
\begin{equation}
W=mcs=\pi \left(\frac{mc^3}{a}\right)=\pi \left(\frac{m^2c^3}{eE}\right).     
\label{cpp8} 
\end{equation}
The Boson weight of such pair production processes summed over the number 
\begin{math} k \end{math} of pairs produced is related to the partition 
function 
\begin{equation}
{\cal Z}=\sum_{k=0}^\infty (-1)^k e^{-kW/\hbar } 
= \left[\frac{1}{1+e^{-W/\hbar }}\right] .     
\label{cpp9} 
\end{equation}
The factor of \begin{math} -1 \end{math} for each semicircle means a 
Bose factor of one for each circle. Since the rate of change of momentum is 
equal to the force, \begin{math} dp/dt = eE  \end{math}, the transition rate 
per unit time per unit length \begin{math} \Gamma_1 \end{math} is given by 
\begin{eqnarray}
\Gamma_1 dt = \left(\frac{dp}{2\pi \hbar }\right)[-\ln {\cal Z}] , 
\nonumber \\ 
\Gamma_1 = \left(\frac{eE}{2\pi \hbar }\right)\ln \left[1+e^{-W/\hbar}\right], 
\nonumber \\ 
\Gamma_1 = \left(\frac{eE}{2\pi \hbar }\right)
\ln \left[1+e^{-(\pi m^2c^3/\hbar eE)}\right],
\nonumber \\ 
\Gamma_1 = \left(\frac{eE}{2\pi \hbar }\right)
\sum_{n=1}^\infty \frac{(-1)^{n+1}}{n}\ e^{-n(\pi m^2c^3/\hbar eE)}.
\label{cpp10}
\end{eqnarray}

By taking the momentum perpendicular to the electric field into 
account, the (3+1) dimensional result follows from Eq.(\ref{cpp10}),   
\begin{eqnarray}
\Gamma =\left(\frac{eE}{2\pi \hbar }\right)\int 
\left[\frac{d^2 {\bf p}_\perp}{(2\pi \hbar)^2}\right]\times 
\nonumber \\     
\sum_{n=1}^\infty \frac{(-1)^{n+1}}{n} 
\exp \left( - \pi n\left|\frac{m^2c^3+cp_\perp ^2}{\hbar e E}\right|\right), 
\nonumber \\ 
\Gamma = \frac{c}{8\pi^3}\left(\frac{eE}{\hbar c}\right)^2 
\sum_{n=1}^\infty \frac{(-1)^{n+1}}{n^2} 
\exp \left( - \pi n\left|\frac{m^2c^3}{\hbar e E}\right|\right),  
\label{cpp11} 
\end{eqnarray}
in agreement with Eq.(\ref{bpp2}). It is not difficult to write the 
transition rate for producing pairs wherein the charged 
particles have spin \begin{math} s \end{math}, i.e. 
\begin{eqnarray}
s=0,\ 1,\ 2,\ 3,\ \cdots \ \ \ \ \ \ \ \ \ \ \ \ \ {\rm (Bosons)},
\nonumber \\ 
s=1/2,\ 3/2,\ 5/2,\  \cdots \ \ \ \ \ \ \ \ \ \ {\rm (Fermions)}. 
\label{cpp12a}
\end{eqnarray}
The statistical index may be defined as 
\begin{eqnarray}
\eta_s =\exp\big[i\pi (2s+1)\big], 
\nonumber \\ 
\eta_s = -1 \ \ \ \ \ \ \ {\rm (Bosons)},
\nonumber \\ 
\eta_s = +1 \ \ \ \ \ \ {\rm (Fermions)}. 
\label{cpp12b}
\end{eqnarray}
From a quantum field theory viewpoint, the statistical index is related 
to the commutation or anti-commutation relation between creation and 
destruction operators
\begin{equation}
[a,a^\dagger ]_{\eta_s}=a a^\dagger +\eta_s a^\dagger a =1
\label{cpp12c}
\end{equation}
For arbitrary spin, Eq.(\ref{cpp11}) may be argued from the factor 
\begin{math} -\eta_s  \end{math} for each closed circle loop to be 
\begin{eqnarray}
\Gamma = \frac{(2s+1)c}{8\pi ^3}\left(\frac{eE}{\hbar c}\right)^2 \times 
\nonumber \\  
\sum_{n=1}^\infty \frac{\eta_s^{(n+1)}}{n^2} 
\exp \left( - \pi n\left|\frac{m^2c^3}{\hbar e E}\right|\right). 
\label{cpp12d}
\end{eqnarray}
Eq.(\ref{cpp12d}) has been discussed in the literature\cite{Bialas:1984}.

Finally, the Euclidean action \begin{math} W \end{math}
may be associated with an entropy \begin{math} S \end{math} via 
\begin{equation}
\frac{W}{\hbar }=\frac{S}{k_B}=
\pi \left(\frac{m^2 c^3}{\hbar eE}\right).    
\label{cpp12} 
\end{equation}
The derivative of the entropy with respect to the rest energy determines 
the reciprocal temperature 
\begin{equation}
\frac{1}{c^2}\left(\frac{dS}{dm}\right) = \frac{1}{T} 
\ \ \ \Rightarrow \ \ \ k_BT = \left(\frac{\hbar eE}{2\pi mc}\right).    
\label{cpp13} 
\end{equation}
In terms of the {\em acceleration} of the charged Bosons, there exists an effective 
temperature\cite{Unruh:1976},  
\begin{equation}
k_BT = \left(\frac{\hbar a}{2\pi c}\right),      
\label{cpp14} 
\end{equation}
of the environment inducing position fluctuations equivalent to the  
energy fluctuations in the rest frame of the applied electric field.

\section{Conclusion \label{conc}}

The central results of this work are not new. For example, the spin zero charged Boson pair 
production rate in Eq.(\ref{bpp2}) as well as its generalization to the general spin $s$
charged particle pair production rate in Eq.(\ref{cpp12d}) are very well known. However, the 
derivations, physical pictures and consequences of zero point oscillations 
are to our knowledge original. The notion of zero point energy in relativistic 
quantum field theories is made real by the particle and anti-particle 
content of the theory. In electromagnetic theory, the photon is its own 
anti-particle which may obscure the interpretation of zero point energy. 
By considering a generic field theory with charged particles, the differences 
in charge between the particle and anti-particle makes manifest the physical 
nature of zero-point oscillations. The vacuum polarization and 
vacuum magnetization can be computed via this simple physical 
viewpoint.

\section*{Acknowledgments}

J. S. would like to thank the United States National Science Foundation for support under PHY-1205845.

\vfill
\eject

\end{document}